**Coalescence 2.0: a multiple branching of recent theoretical developments and their applications.**

Aurélien TELLIER[1]

Christophe LEMAIRE[2]

[1] *Section of Population Genetics, Center of Life and Food Sciences Weihenstephan, Technische Universität München, 85354 Freising, Germany*

[2] *Institut de Recherche sur l'Horticulture et les Semences, University of Angers, Angers, France*

Corresponding author:

Aurélien TELLIER

*Section of Population Genetics, Center of Life and Food Sciences Weihenstephan, Technische Universität München, 85354 Freising, Germany*

Email: tellier@wzw.tum.de

ABSTRACT

Population genetics theory has laid the foundations for genomics analyses including the recent burst in genome scans for selection and statistical inference of past demographic events in many prokaryote, animal and plant species. Identifying SNPs under natural selection and underpinning species adaptation relies on disentangling the respective contribution of random processes (mutation, drift, migration) from that of selection on nucleotide variability. Most theory and statistical tests have been developed using the Kingman's coalescent theory based on the Wright-Fisher population model. However, these theoretical models rely on biological and life-history assumptions which may be violated in many prokaryote, fungal, animal or plant species. Recent theoretical developments of the so called multiple merger coalescent models are reviewed here ($\Lambda$-coalescent, beta-coalescent, Bolthausen-Snitzman, $\Xi$-coalescent). We explicit how these new models take into account various pervasive ecological and biological characteristics, life history traits or life cycles which were not accounted in previous theories such as 1) the skew in offspring production typical of marine species, 2) fast adapting microparasites (virus, bacteria and fungi) exhibiting large variation in population sizes during epidemics, 3) the peculiar life cycles of fungi and bacteria alternating sexual and asexual cycles, and 4) the high rates of extinction-recolonization in spatially structured populations. We finally discuss the relevance of multiple merger models for the detection of SNPs under selection in these species, for population genomics of very large sample size and advocate to potentially examine the conclusion of previous population genetics studies.

**Introduction**

Since the end of the 20th century, and increasingly recently, molecular data are being used to reveal the evolutionary history of populations. Population genetics and genomics approaches provide answer to key evolutionary questions such as understanding which evolutionary forces drive genome evolution, or pinpointing the molecular bases for species or population adaptation to their environment (biotic or abiotic). Population genetics is firmly grounded on the mathematical theory founded by the seminal work of Wright (Wright 1931), Fisher (Fisher 1930), Malécot (Malecot 1941) and Kimura (Kimura 1954) amongst others, stating that genomes evolve by the action of random neutral processes (mutation, drift and migration) and natural selection (positive or negative). In this respect, the so-called Kingman coalescent (Kingman 1982) based on the Wright-Fisher (Fisher 1930; Wright 1931) and Moran (Moran 1958) models (see description below) has been instrumental for connecting mathematical and stochastic theory with polymorphism data. The Kingman coalescent allows us to reconstruct the genealogy of a sample of individuals from present (and past) DNA polymorphisms, opening the possibility of model-based statistical inference (Rosenberg & Nordborg 2002).

An important result of population genetics (Kingman 1982; Wakeley 2008) is that neutral random processes, such as past demographic expansion, can generate similar patterns of nucleotidic variability in the genome as those resulting from natural selection, such as positive selection (Tajima 1989). However, it is also assumed that demographic events affect the whole genome, whereas selection affects potentially only few loci. All genomics studies using statistical methods to detect natural selection, which are based on polymorphism data (Tajima's D, Fay and Wu's H, McDonald-Kreitman) or drawing inference of past demography and/or selection rely extensively on the predictions from theoretical population genetics, sometimes more specifically from the coalescent theory. This theoretical framework and developed statistical inference methods have been extensively used to study demography of species and populations (Nelson *et al.* 2012), recent speciation events (review in Sousa & Hey 2013), existence of seed banks (Tellier *et al.* 2011), and the occurrence of natural selection (Hernandez *et al.* 2011). Population genetics studies have been conducted using the Wright-Fisher and Kingman coalescent framework on a wide range of organisms, ranging from mammals to bacteria, fungi or plants which vary greatly in their generation time, population sizes, overlapping of generations, mutation rates, spatial structures, ecological characteristics and population dynamics. However many of the studied species or populations violate the major underlying model assumptions of the Wright-Fisher and Kingman coalescent *i.e.* panmixia, constant population size, small variance in offspring number and coalescence of two lineages at a time. Even though much theoretical work has been done on extending these models contributing to our understanding of the evolutionary forces shaping patterns of polymorphism at the population or species level, it is, however, not always clear to what extent these violations affect statistical inferences from polymorphism data.

It becomes of interest to investigate recent theoretical developments as a source for improving our current genomics and model based inference methods. Starting with the Bolthausen-Snitzman coalescent (Bolthausen & Sznitman 1998), followed by the $\Lambda$-coalescent (Donnelly & Kurtz 1999; Pitman 1999; Sagitov 1999) and $\Xi$-coalescent (Schweinsberg 2000; Möhle & Sagitov 2001) there is a recent burst of mathematical theory, which aims to generalize the Kingman model (Wakeley 2013). In a nutshell, these theories allow for the merging (coalescence) of multiple lineages (more than two) at a given generation, and possibly several simultaneous mergers (Fig. 1). They are thus referred thereafter as Multiple Merger Coalescent (MMC) models. A common theme in these studies is the demonstration that the Kingman coalescent is a peculiar case of the general class of MMC models. As such, these models may present a greater range of applicability for genomic analysis of species with peculiar life cycles, which infringe the assumptions of the Wright-Fisher or Moran models and Kingman coalescent.

After summarizing the main assumptions and limitations of the current coalescent theory based on the Wright-Fisher model, we turn our attention to the MMC models. We have five major aims: 1) to provide the first overview of this burgeoning literature on MMC for a non-mathematical audience, 2) to analyse the assumptions and limitations of these models in comparison to the Kingman coalescent, 3) to review the current applications of MMC models for data analysis, 4) to advocate that MMC models are especially suitable for coalescent studies in a wide range of species (plant, animal, fungi, bacteria, viruses) which exhibit peculiar life cycle or life history, and 5) to highlight the need to use MMC models in such species to improve the statistical analysis of genomic data.

**The Kingman coalescent: assumptions, extensions and limitations**

The Kingman coalescent (Kingman 1982) is obtained as a backward in time continuous limit of the discrete Wright-Fisher (Fisher 1930; Wright 1931) or Moran (Moran 1958) models. The Wright-Fisher model assumes no overlapping generations and a panmictic population of $2N$ haploid individuals, or $N$ diploid individuals. All individuals reproduce by producing gametes and then die at each generation $t$. Panmixia and constant population size are obtained assuming that at the generation $t+1$, each offspring individual picks one ancestor at random in the parental generation $t$. Two properties of the Wright-Fisher model are crucial to derive the Kingman coalescent: the large population size $N$ compared to the sampling size $n$ from which polymorphism data are obtained, and the small and finite variance of offspring number per parental individual.

Assuming a sample size of $n$ individuals at present, the Kingman coalescent is obtained as a limiting genealogy when the population size $N$ is sufficiently large, especially compared to the sample size $n$ ($N \to \infty$ and $n \ll N$ in mathematical terms). The genealogy can be traced back from a sample size of $n$ individuals, to a most recent common ancestor (MRCA) due to the finite size of the population and the effect of genetic drift (*e.g.* Kingman 1982; Wakeley 2008). The second important condition for the coalescent approximations to hold, is that the parental individuals produce a small number of offsprings when compared to the population size $N$. In the Wright-Fisher model, the offspring distribution per parental individual follows a binomial distribution with mean 1, and variance (1-1/2$N$), which is approximately equivalent to a Poisson distribution with mean 1 and variance 1. A coalescent event is defined as the merging of $k$ lineages at some point in time (with $k = 2$). The two key assumptions above assure that both the probability that more than two lineages coalesce at once (a so-called multiple merger event) and that of simultaneous events to occur are negligible (on the order of $O(1/N^2)$ when $N$ is large enough). In mathematics, $O(1/x^2)$ would capture all the terms that decrease to zero at rate $1/x^2$ or faster, as $x$ becomes very large (goes to infinity, $x \to \infty$). The property of a coalescent tree, *i.e.* its topology and the size distribution of all branches, is thus defined by the frequency of merging of lineages (the coalescent rate) and how many lineages coalesce, with time scaled by the population size $N$. To obtain possible polymorphism data under a given model, mutations can be thrown on the genealogy following a Poisson process with the population mutation rate $\theta = 4N\mu$ where $\mu$ is the mutation rate per base pair per generation (description in Wakeley 2008). Note that in the Wright-Fisher model, the effective population size (*Ne*) is equal to the observable population size (*N*).

The Kingman coalescent has been instrumental in analysing polymorphism data because it can be easily modified to relax the assumptions of constant population size in time (*e.g.* (Watterson 1984; Kaj & Krone 2003), single panmictic population (*e.g.* Wakeley & Aliacar 2001; Charlesworth *et al.* 2003), and non-overlapping generations (Tellier *et al.* 2011). Modifications of the Kingman coalescent are obtained by using time-rescaling argument to accommodate variable rates of coalescence in time (Kaj & Krone 2003). For example, an increase in population size is accommodated by shortening coalescent times and increasing the coalescent rates, assuming that all individuals present a similar increase in their per capita offspring production, and that the average offspring numbers is still much smaller than *N*. Analogues of the Kingman coalescent were also built to accommodate sexual reproduction and intra-locus recombination (the Ancestral Recombination Graph, Hudson 1983) and natural selection (the Ancestral Selection Graph, Krone & Neuhauser 1997). We redirect interested readers to in depth reviews about the Kingman coalescent (Wakeley 2008), and the use of coalescent simulators (Hoban *et al.* 2012).

However, recent studies in various marine organisms (sardines, cods, oysters…) have suggested that some individuals produce a number of surviving offspring on the order of magnitude of *N*. Indeed high fecundities but also high early mortality characterise most of the marine organisms. This effect independent of natural selection and entirely driven by a high neutral variance of reproductive success, is named the "sweepstake reproduction" (Beckenbach 1994; Hedgecock 1994; Li & Hedgecock 1998; Hedgecock & Pudovkin 2011). One major consequence is that populations experience successions of strong random genetic drift followed by demographic expansions of few lineages. It has been also noted as well that under the action of natural selection, favoured individuals may produce many offsprings (on the order of *N*) over short period of time (Schweinsberg & Durrett 2005; Coop & Ralph 2012). In these cases, the classical assumption of finite, and small compared to *N*, variance in offspring production may be violated in many animal, plant, bacterial or fungal species. This is the starting point for the development of MMC models.

**Overview of existing MMC models**

Our objective here is to describe the general framework of multiple merger coalescent models and not their rigorous mathematical description. We redirect the interested reader to the cited literature for rigorous mathematical definitions (see Berestycki 2009; Etheridge 2011). MMC models are derived from the general Cannings model of population dynamics (Cannings 1974; Sagitov 1999; Möhle & Sagitov 2001), from which the Moran and Wright-Fisher models are specific cases. We describe here the recent model of Schweinsberg (2003) which has been used to derive the $\Lambda$-coalescent, but which idea captures the essence of the population model underlying the MMC models. In a population of size *N* each haploid individual independently produces a random number of juvenile offsprings following a given probability distribution, allowing for large number of offsprings. At each generation, density-dependent regulation operates in the population so that exactly *N* juveniles, sampled at random, will survive to maturity and constitute the next generation. The mean number of juveniles produced by each parental individual is thus assumed to be greater than one and there are thus always many more than *N* juveniles to choose from. MMC models deal thus with the case where individual offspring distributions have a large variance, *i.e.* the number of chosen juveniles from a given individual parent can approach *N* with a non-negligible probability. Note however, that after population regulation the mean number of offspring per parental individual is assumed to be one, because population size is constant in time as in the Wright-Fisher or Moran models.

The most general model of multiple merger coalescent is the so-called $\Xi$-coalescent in which simultaneous multiple mergers of lineages are possible, or in other words several collisions of *k* lineages ($k \geq 2$; Fig. 1). This class of model was introduced by Möhle & Sagitov (2001) and described in its full generality by Schweinsberg (2000). We mention this class only briefly in this review because it is less studied for applications to polymorphism data analysis than the following $\Lambda$-coalescent models (but see *e.g.* Birkner *et al.* 2009; Taylor & Véber 2009).

A second class of model is the $\Lambda$-coalescent, which allows only one merging of multiple *k* lineages ($k \geq 2$) at any point in time (Fig. 1). This class of models was established independently by Donnelly & Kurtz (1999), Pitman (1999), and Sagitov (1999). It specifies the genealogy of a sample from a Cannings population model based on the so-called $\Lambda$-Fleming-Viot process (Bertoin & Gall 2003). Based on the definition of the rate of coalescence (see Appendix), the Kingman's coalescent is a special case of Λ-coalescent where only two lineages are allowed to merge at a time ($k=2$). The Λ-coalescent can be defined by several distributions of the frequency of multiple mergers, that is how often multiple coalescent events occur, and the size of merging events, that is the number of coalescent lineages $k$. Recently, properties of Λ-coalescent genealogies have been studied such as the expected site-frequency spectrum (SFS) and number of segregating sites (Berestycki et al. 2014; Birkner et al. 2011; 2013).

The Beta-coalescent is a sub-class of Λ-coalescent models which were obtained by Schweinsberg (2003). It is defined as the Beta($\alpha$, 2-$\alpha$) coalescent with a specific rate of multiple coalescent events α (see Appendix). In mathematical terms, multiple merger events occur on the time scale of O($1/N^{\alpha}$) (with $\alpha < 2$) as described in Birkner and Blath (2008). The Beta-coalescent is obtained as a specific case of Λ-coalescent assuming a beta distribution, with parameter α, of multiple merger probabilities (Schweinsberg 2003). The properties of the beta-coalescent are well studied and various aspects of its genealogies computed such as length of coalescent trees, the expected site-frequency spectrum (SFS) and number of segregating sites (Fig. 2, (Berestycki *et al.* 2007, 2008; Birkner & Blath 2008). These results are key for the potential application to analysis of polymorphism data. For example, in a Kingman coalescent, the ratio $R_1 = T_1/T$, where $T_1$ is the external branch lengths of a coalescent tree and $T$ is the total branch lengths of a tree, has been show to be equivalent to $R_1 \approx \xi_1/S$ where $\xi_1$ is the number of segregating sites at frequency one (singletons) and $S$ is the total number of segregating sites. For the Beta-coalescent, the ratio $R_1$ tends to the value 2-α for large sample size ($n\rightarrow\infty$; Berestycki *et al.* 2007, 2008). In practice, inference of α is possible based on the SFS and the total number of segregating sites of a large enough sample $n$ of individuals (Birkner and Blath 2008; Birkner et al. 2011; Steinrücken et al. 2013).

A peculiar case of Beta-coalescent is the so-called Bolthausen-Snitzman model, though it originates from spin glasses models in physics (Bolthausen & Sznitman 1998). This model is obtained assuming a uniform distribution for the probability of multiple merger (see Appendix) and defined as a beta-coalescent with α=1. Note that this definition implies that the ratio $R_1$ tend in this model to one (when $n\rightarrow\infty$). Therefore, the ratio $R_1$ would not be a useful quantity to calculate from sequence data in this case, as it does not correlate with the model parameter. Expected statistics on the genealogies have also been derived (Basdevant & Goldschmidt 2008). This model was found to reflect the genealogy of models with population under rapid positive selection (Brunet *et al.* 2007; Brunet & Derrida 2012; Neher & Hallatschek 2013; Neher *et al.* 2013).

A final type of model, the ψ-coalescent, was derived by Eldon and Wakeley more recently (Eldon & Wakeley 2006, 2008, 2009; Eldon & Degnan 2012). This model has for starting point the biological assumptions of sweepstake reproduction. The parameter ψ defines the proportion of offspring in the population, which originate from one parent at the previous generation (see Appendix). This model does not present all mathematical properties of a Λ-coalescent (Der *et al.* 2012), but its behaviour allows only one at a time multiple merger of $k$ lineages ($k \geq 2$; Fig. 1). It has been used to infer the strength and occurrence of sweepstake events in semelparous fishes and marine organisms and is thus influential for applications of MMC models (Eldon 2009, 2011). Note nevertheless that the ratio $R_1$ also tends to one (when $n \rightarrow \infty$) in this ψ-coalescent model (Eldon 2009, 2011).

**Patterns of polymorphism under MMC**

In the following, we describe and highlight the main differing features of MMC models in terms of genetic diversity, patterns of polymorphism, linkage disequilibrium and population differentiation, compared to expectations under the classic Kingman coalescent. Furthermore, we discuss for various effects of MMC on genealogies, how the given polymorphism pattern can lead to bias and erroneous conclusions if such pattern is interpreted under the classic Wright-Fisher assumptions.

For the multiple merger coalescent genealogies to exhibit differences from those under the Kingman model, MMC events must be frequent enough. However two extreme situations are predicted. On the one hand, when the rate of multiple merging coalescent events ($k \geq 2$) is much smaller than the rate of binary coalescence ($k =2$), the coalescent process approaches the Kingman model, and no signature of MMC would be observable. On the other hand, if the rate of multiple merging is much higher than the rate of binary coalescence, the amount of genetic diversity is greatly reduced (Eldon & Wakeley 2006, 2008). This is easily understandable from the biological point of view (see Arnason 2004). If the sweepstake events are very seldom and of small size, *i.e.* most individuals in the population reproduce and leave one viable juvenile at most generation, the population's genealogy can be reconstructed by a classic Wright-Fisher model. Conversely, if at almost every generation, one or very few individuals produce all surviving juvenile offsprings, the population will be composed of very few genotypes. These genotypes are very similar from each other because they share a recent common ancestor, and there is very little time for new mutations to appear. A consequence of the sweepstake reproduction mode is then that the effective population size ($Ne$) can be significantly smaller than the total observable population size ($N$; Nunney 1995). Mathematically, this is described as follows: in a Kingman coalescent, $Ne$ scales linearly with $N$, whereas for MMC models, $Ne$ can be a fraction of $N$ or scales on the order of $\log(N)$(Huillet & Möhle 2011). In other words, it is expected that under MMC models, the effective population size measured from polymorphism data yields significantly smaller values than the number of observable individuals $N$ in the population or the species (see below for example). In contrast, the observation of small $Ne$ compared to observable $N$ is yet often currently interpreted under the Wright-Fisher model as a signal of population extinction or strong bias in sex-ratio. We suggest here that these conclusions may be incorrect when studying species undergoing regular sweepstake reproduction with constant population size and balanced sex-ratio (see Hedgecock 1994; Arnason 2004).

*Effect on genetic diversity and the site-frequency-spectrum*

A further noticeable characteristic of MMC models is that the coalescent process occurs on a faster time scale than under the Kingman model. Indeed coalescent trees obtained under MMC models exhibit more star like shaped genealogies and skews in the resulting SFS with an excess of low frequency variants (*e.g.* singletons, Fig. 2 blue bars) generating a negative Tajima's D (Birkner *et al.* 2013) than under the Kingman model with constant population size (Fig. 2 black bars). The star like phylogeny and skew in SFS will be more pronounced with increasing rates of multiple merging, for example for smaller values of α (with α < 2) in the Beta-coalescent model or higher values of $\psi$ (the proportion of offspring derived from one individual in the population) in the $\psi$–coalescent (Fig. 2). MMC models with constant population size generate in fact an excess in length of external branches in the genealogies, which can be compared to that arising from Kingman coalescent model with strong recent population expansion (Fig. 2 red bars, Birkner et al. 2013; Steinrücken et al. 2013). The usual interpretation that lack of genetic diversity and excess of low frequency SNPs (and negative Tajima's D) is a signal for past bottleneck or at least very strong population expansion (Fig. 2), may be often incorrect, when studying species undergoing sweepstake reproduction.

*Effect on genetic drift and linkage disequilibrium*

The rate of genetic drift (Der *et al.* 2011, 2012) and the amount of linkage disequilibrium (Eldon & Wakeley 2008) are also affected in MMC models compared to the Kingman expectations. Assuming that the rate at which MMC events occur is high enough, *i.e.* on the order of the coalescent rate, show that the strength of genetic drift decreases in MMC models compared to the Wright-Fisher model (Der *et al.* 2011, 2012). In a Wright-Fisher model with a bi-allelic system (two alleles *a* and *A*), at every generation each offspring chooses randomly a parent with a probability equal to that of each allele frequency. The frequencies of *a* and *A* are then given by binomial sampling at every generation. Such a case represents an upper limit to the strength and occurrence of drift (Der *et al.* 2011). In a model where both alleles *a* and *A* are neutral and their frequencies are only driven by drift, MMC models generate long period of frequency stasis (unchanged values) only interrupted by multiple merger events. The fixation probability of a new allele is unchanged in MMC models compared to a Wright-Fisher model and is equal to its initial frequency (namely $1/2N$; Der *et al.* 2011). However the time to fixation of a mutant allele entering the population at frequency $1/2N$ decreases from $2N$ in the Wright-Fisher model to $N \times \log(N)$ in MMC models.

MMC models generate more pronounced star like genealogies (see above). This has for consequences that events affecting the genealogy, such as intra-locus recombination, can affect a smaller or larger number of branches compared to that expected under the Kingman's coalescent (Eldon & Wakeley 2008; Birkner *et al.* 2012). In biological terms, this means that if sweepstake reproduction events are frequent, the efficiency of recombination and meiosis in reshuffling genotypes is very limited as present genotypes descent from a very recent ancestor (Eldon & Wakeley 2008). As a result under MMC models, the amount of linkage disequilibrium (LD) can be uncorrelated to the genomic recombination rate (the Ancestral Recombination Graph for MMC model; Birkner *et al.* 2012). For example, high LD can be observed despite high genomic rates of recombination and *vice and versa*, depending on where multiple merging events occur in the coalescent tree (Eldon & Wakeley 2008). Moreover, genealogies for loci far apart on the same chromosome may remain correlated, and LD is a function of the rate of recombination and of the reproduction parameter (of the skew in offspring distribution; Birkner *et al.* 2012). These counter-intuitive results suggest that for species undergoing sweepstake reproduction, recombination hot-spots may not be detectable by measuring recombination rates based on SNP frequencies. Conversely, the analysis of such species under the paradigm of Kingman model, would lead to over or under-estimate the rates of recombination, and potentially misestimate the adaptive potential of a given species (Eldon & Wakeley 2008). Additionally the misestimate of recombination rates would strongly affect the outcomes of analysis of hitchhiking in genomic islands of differentiation. Indeed as the width of hitchhiked zones depends mainly of recombination and selection parameters (Barton 2000), effects of MMC on the shape of genomic islands should be taken into account.

*Effect on Fst and measure of population differentiation*

MMC models also affect the shape of genealogies of a sample in models with several populations. A key factor to consider when comparing MMC and Kingman models, is that the time and effective population size would scale differently because of the different coalescent rates (e.g. Donnelly & Kurtz 1999; Pitman 1999; Sagitov 1999; Eldon & Wakeley 2009). In a model with several population linked by migration, $F_{ST}$ is a commonly computed index of allelic fixation. In the Wright-Fisher model, for a given mutation rate, $F_{ST}$ is a function of the product $N \times m$ where $N$ is the population size and $m$ is the migration rate. Due to the difference in time scaling in Kingman and MMC models, (Eldon & Wakeley 2009) show that a similar value of $F_{ST}$ can be obtained for a given $N^2 \times m$ in a Moran model, and for $N^\gamma \times m^*$ in a MMC model, with $m^* > m$. (Note that following a time-rescaling argument, migration depends on $N^2$ in the Moran model). The scaling parameter of the MMC model is $\gamma$, regulating the frequency of large offspring production ($0 < \gamma < 2$) and for which high values tend to produce Kingman coalescent genealogies, and $m^*$ is the migration rate (in a $\psi$-coalescent, but also valid for a Beta-coalescent; Eldon & Wakeley 2009). In biological terms, this means that in spatially structured populations of a species with sweepstake reproduction, interpreting values of $F_{ST}$ as under the Kingman coalescent would yield a clear underestimation of both *N* and the migration rate (*m*). In other words, for a given identical scaled migration and *N* in both MMC and Kingman models, higher rates of alleles fixation in each population and thus higher $F_{ST}$ are expected to occur in MMC model.

In models of recent speciation where two incipient populations split from an ancestral one without post-divergence migration (Wakeley & Hey 1997), Eldon & Degnan (2012) show for a Λ-coalescent and ψ-coalescent models that the probabilities of monophyly or paraphyly of each population vary compared to the Kingman expectations. The number of incomplete lineage sorting, which is used by several methods to determine the structure of populations and times of split, is affected by MMC. In practice, this means that two species with sweepstake reproduction, which have split at time *τ* ago, would be estimated to have an older time of split and more incomplete lineage sorting when analysed under the classic assumptions of the Kingman setting (e.g. Wakeley & Hey 1997; Sousa & Hey 2013). This is because the time of allele fixation in MMC is shorter, so that more divergence (and substitutions) accumulates between the two incipient populations without gene flow than expected under the Kingman model.

*Effect on natural selection*

As MMC models affect the rate and strength of genetic drift, Der *et al.* (2011, 2012) demonstrate that the time of fixation and probability of fixation of alleles under positive or negative selection differ from the classic Wright-Fisher expectations. Sweepstake reproduction can either suppress or amplify the strength of selection (Der *et al.* 2011). A surprising result is for example that alleles under positive selection may have a probability of one to become fixed under MMC models, where it is of 2*s* in a classic Wright-Fisher model (where *s* is the coefficient of selection; Kimura 1954). In other words, as drift becomes weaker under MMC models, the efficacy of selection increases, because selection acts almost in a deterministic manner on allele frequencies during the phases between multiple merger events. Practically, these results point out two interesting features and potential bias in our current interpretation of population genetics results. 1) Natural selection coefficients estimated under Wright-Fisher model are over-estimates of the real coefficients for population undergoing sweepstake reproduction. 2) The efficacy of positive selection to favour alleles or that of negative selection to remove deleterious ones, is greatly underestimated when ignoring the sweepstake reproduction in a population.

**Current and potential applications of MMC**

We present here a series of ecological set-ups, life history traits, species and biological systems for which genomic evolution in populations is likely better pictured by MMC rather than the currently used Kingman and Wright-Fisher models. We first describe neutral and then selective mechanisms which generate MMC genealogies. It is important to distinguish between these two types of mechanisms as neutral random processes affect the dynamics of diversity over the whole genome, whereas selective processes generate MMC at few loci or genomic regions.

*Neutral mechanism: Skew in offspring production per capita*

Sweepstake reproduction defined as the high variance of reproductive success amongst individuals at every generation by random effect and density-dependent regulation of population has been suggested early on as a key characteristic of marine species (Beckenbach 1994; Hedgecock 1994; Fig. 3A and the model of Schweinsberg 2003). This has led to the development of empirical studies using genotyping in fish and crustaceans to confirm this effect for example in Pacific Oysters (Boom *et al.* 1994; Boudry *et al.* 2002) or atlantic cod (Li & Hedgecock 1998; Arnason *et al.* 2000; Arnason 2004). These studies have been recently reviewed in (Hedgecock & Pudovkin 2011). This group of organisms and genotyping data has been instrumental in the development of the statistical application of MMC models. The aim is here to infer from SNP data the rate of sweepstake events, that is how often they occur and which percentage of the population is affected. Likelihood methods have thus been developed for the $\Lambda$-coalescent (Birkner et al. 2011), Beta-coalescent (Birkner and Blath 2008; Steinrücken et al. 2013) or $\psi$-coalescent (Eldon & Wakeley 2006; Cenik & Wakeley 2010; Eldon 2011) in order to infer the parameters of interest based on the site frequency-spectrum.

A population with strong density dependent size regulation between juveniles and adults is included in (Schweinsberg 2003) or in Eldon and Wakeley ψ-coalescent models. These are based on the classic sweepstake hypothesis. However, this mechanism, and these models, may also apply generally to plant populations where the production of seeds exceeds largely the carrying capacity of a population. This is suggested for trees (Ingvarson 2010), and it may be a general feature of annual plants with fast growth colonizing disturbed habitats and producing numerous seeds (such as *Arabidopsis thaliana*). This model can also be extended to many species like insects that experience several population boosts and bursts (Wallner 1987). The suitability of MMC models to explain patterns of diversity can be tested using genomic data in numerous plant populations (*e.g.* in *Arabidopsis thaliana*) using the existing likelihood methods (Birkner and Blath 2008; Steinrücken et al. 2013). The applicability of MMC models for given plant and marine species has two impacts. First, this may drive us to reconsider predictions and analyses of polymorphism patterns in response to local adaptation and natural selection (*e.g.*Savolainen *et al.* 2013). Second, it would impact conservation biology measures to be adopted as the population size may not be a good indicator of future persistence if sweepstake events occur and can deplete the population from its genetic diversity (Hedgecock & Pudovkin 2011).

*Neutral mechanism: Parasitic life cycle*

Parasites and pathogens such as viruses, fungi, oomycetes or bacteria may often present several life-history traits and life cycles which promote MMC genealogies such as 1) sweepstake reproduction, 2) recurrent bottlenecks between growing seasons or during between host transmission, and 3) alternate asexual and sexual phases.

Firstly, pathogens, in particular plant or insects parasites producing aerial infectious spores, may present typical sweepstake reproduction. Spores production is well studied in crop pathogens, which can generate several thousands or millions of spores per lesion (Agrios 2005), as shown for two widely spread pathogen of wheat *Zymoseptoria trictici* (septoria of wheat, Simon & Cordo 1997) and *Puccinia striiformis* f. sp. *tritici* (wheat yellow rust, Hovmøller *et al.* 2011). However when measuring effective population sizes using DNA polymorphism data, population sizes on the order of only few thousands are typically found at the continental scale: 5,000 for *Z. tritici* (Stukenbrock et al. 2011;

2012) and 25,000 for yellow rust (Duan *et al.* 2009). Many plant pathogens may thus fulfil the criteria of sweepstake reproduction and of a large difference between observable population size and effective size (Fig. 3A), possibly due to a large majority of spores being unable to find available hosts in many geographic areas (the oasis in the desert metaphor, Brown et al. 2002).

Secondly, many parasites exhibit recurrent strong variation in population sizes. Strong recurrent bottlenecks will affect the whole genome, and the underlying genealogies are shown to converge to a Ξ-coalescent (Birkner *et al.* 2009). A first type of bottleneck occurs due to limited host availability. Crop pathogens in particular exhibit reduced population sizes at the end of the growing season when hosts are not available post-harvest (Fig. 3B; Stukenbrock & McDonald 2008). Similarly, host populations in the wild (insects or plants) are often unavailable or very drastically reduced during the winter (unfavourable) season. A second type of bottlenecks which is common to many parasites occurs during between host transmissions. Animal (e.g. malaria, virus) or plant (viruses or bacteria) parasites which are transmitted by vectors (aphids, mosquitoes, …) experience strong bottlenecks as few viral particules or bacteria are stochastically transmitted to the next host (Fig. 3C; Moury *et al.* 2007; Pybus & Rambaut 2009). For a review and meta-analysis see the recent study by Gutiérrez *et al.* (2012). A third type of bottleneck is produced inherently by epidemiological disease dynamics over time, where parasite population sizes show regular expansions and contractions (Pybus & Rambaut 2009). Balloux & Lehmann (2012) show that the interaction of varying population sizes due to strong recurrent bottlenecks with overlapping generations generates a neutral increase of substitution rates in the genome. This case applies particularly to many human parasites such as influenza or plague (Morelli *et al.* 2010), but potentially to several crop and wild plant pathogens. The result is that variable mutation rates along branches of a genealogy of various parasite strains may be generated by these neutral random stochastic processes as emergent properties of MMC models, and calls for caution when inferring signatures of selection in the genome (for example using dN/dS ratio).

Thirdly, the life-cycle of many fungal plant pathogens (rusts, mildew, oomycetes), but also some plant species, crustaceans (Daphnia) or insects (aphids) comprising alternating phases of clonal reproduction and one or few sexual events per year, is expected to generate typical MMC genealogies. This occurs as the clonal phases generate a large amount of offsprings with a large reproductive variance among genotypes, with few clones possibly increasing in frequency due to random processes and clonal interference (Neher 2013). Moreover, the sexual phase which occurs rarely and only depending on the environmental conditions, may represent a recurrent bottleneck for the survival of individuals to the next generation. Some rust pathogens for example reproduce sexually on a secondary aecidian host with a bottleneck possibly occurring at this stage (Duan *et al.* 2009). Understanding MMC models generated under such life-cycles is of importance for predicting plant, animal, pests and pathogen adaptation in space and time, and the emergence of aggressive or virulent clonal lineages (Stukenbrock & McDonald 2008; Pybus & Rambaut 2009).

*Neutral mechanism: Spatial structure with extinction/recolonization*

Most species live as a metapopulation which consists of demes, *i.e.* local panmictic populations, connected by migration and subjected to regular extinction and recolonization events (Hanski 1998). This is a feature of many plant species (Freckleton & Watkinson 2002) and crustaceans such as *Daphnia* living in ephemeral habitats (Haag *et al.* 2005). The Kingman coalescent applied to a metapopulation suggests that the genealogy is divided into a short scattering phase and a long collecting phase (Wakeley & Aliacar 2001), while the rates of coalescence in these phases depend on the deme size and level of gene flow between demes. Using different sampling strategies, it is possible to take into account the genealogies of these phases and to study neutral and selective processes acting at the local (scattering phase) and the whole species (collecting phase) level (Städler *et al.* 2009; Tellier *et al.* 2011).

However, in a model with regular extinction and recolonization and a very large number of demes, the genealogy can present locally simultaneous multiple mergers (Ξ-coalescent; Limic and Sturm 2006; Taylor & Véber 2009; Barton *et al.* 2010). MMC genealogies are thus expected over the whole genome of individuals in each deme, and to affect the study of evolutionary processes locally, that is the inference of demography and selection in the scattering phase. However, in a model assuming a Λ-coalescent in each deme, the genealogy underlying a species wide sampling of individuals (Städler *et al.* 2009), that is reflecting the collecting phase, tends to converge to the classic Kingman's coalescent (Heuer & Sturm 2013). It follows that the collecting phase may be well approximated by the classic

Kingman coalescent, even though MMC genealogies occur locally within each deme. The occurrence of Λ- or Ξ-coalescents within structured populations following different population dynamics, especially strong extinction-recolonization, and sampling schemes has thus consequences for our understanding of polymorphism data (*e.g.* measure of $F_{ST}$) and for correctly estimating gene flow among populations (see above). Using an ad hoc model to represent the genealogies may be important in conservation biology for studying species in fragmented habitats and the consequences of the decrease or increase in gene flow on the genetic diversity. Additionally, MMC models may be relevant to study the population structure of crop pathogens which undergo long distance dispersal and regular extinction-recolonization such as wheat yellow rust (*P. striiformis* f. sp. *tritici*; Brown and Hovmøller 2002).

*Neutral mechanism: When the sample size is very large*

MMC models appear under another violation of the Kingman assumptions, namely when the sample size (*n*) is larger or on the same order of magnitude than the effective population size (*N*). Wakeley & Takahashi (2003) demonstrate that when $n \gg N$, there is a short phase of numerous simultaneous multiple coalescent events, much like a Ξ-coalescent model. Interestingly, using such large sample size, it is possible to estimate the population size (*N*) and the mutation rate separately, because the number of singletons is then an estimator of *n*/*N* (Wakeley & Takahashi 2003). This result has been recently used in the human population to estimate the actual European effective population size to be around three millions and an ancestral size of around 7,700 (Coventry *et al.* 2010; Nelson *et al.* 2012). We suggest here that this method can be used for other plant or animal species where the effective population size estimated from sequence data is small (for example if the species is endangered), and sampling of numerous individuals is feasible.

*Selective mechanisms: Positive selection and sweeps*

MMC models are also shown to represent genealogies at loci under positive selection (Durrett & Schweinsberg 2005; Schweinsberg & Durrett 2005). Selective sweeps, generate the so-called hitch-hiking effect around the selected site (Maynard-Smith & Haigh 1974). Genome scans are used extensively to detect loci underlying species adaptation to their environment, based on the principle that these few loci show an outlier genealogy compared to the rest of the genome. Durrett & Schweinsberg (2005) show that genealogies during and shortly after selective sweeps are well-approximated by certain Ξ-coalescents. This is not surprising as positive selection creates an excess of low and high frequency variants due to the high rates of coalescence observed during the selection phase (Coop & Ralph 2012). Detecting recent positive selection at a given locus is thus equivalent to estimate if its genealogy fits better to a MMC model rather than the classic Kingman compared to other loci in the genome. Existing likelihood methods for MMC models may be used for genome-wide scans for loci under positive selection.

*Selective mechanisms: Parasitic life cycle and fast adaptation*

A final important occurrence of MMC models is observed in parasites which undergo several multiplications per host generations, especially via long period of intra-host evolution, and are subjected to strong within host selection. This intra-host evolution in facultative asexual parasites is driven by strong genetic draft, due to clonal interference and continuous positive selection as shown in Fig. 3C (Neher & Shraiman 2011; Desai *et al.* 2013; reviewed in Neher 2013). Viruses for example undergo strong bottlenecks during between host transmissions and exhibit huge subsequent diversification driven by intra-host selective processes (Fig. 3C; Moury *et al.* 2007; Pybus & Rambaut 2009; Neher & Hallatschek 2013). The model of selection for fast adapting parasites clearly violates the classical models based on the Kingman coalescent and is shown to follow a Bolthausen-Snitzman coalescent model (Neher *et al.* 2013). The genealogy is driven by strong directional selection with continuous adaptation for a given fitness trait (Brunet *et al.* 2007; Brunet & Derrida 2012; Neher & Hallatschek 2013). The Bolthausen-Snitzman coalescent model has been applied to polymorphism data from human viruses (Neher & Hallatschek 2013; Neher *et al.* 2013), but may be in principle relevant to polycyclic crop viruses, bacteria and fungi (for example the aforementioned *Z. tritici* and *P. striiformis* f. sp. *tritici*, but also other rusts or oomycetes).

Note also that the strong efficacy of selection under MMC models (see above, Der *et al.* 2011) and the strong selective pressure exerted by the use of uniform crop genotypes within fields (Stukenbrock &

McDonald 2008) may explain in part the abundance of signatures of positive selection and the pervasive purifying selection found in population polymorphism data of various crop pathogens such as *Z. tritici* (Stukenbrock *et al.* 2011, 2012; Brunner *et al.* 2013) and other species (Wicker *et al.* 2013). The proportion of the genome which is affected by genetic draft due to selection (Neher & Hallatschek 2013; Neher *et al.* 2013), and thus the extent of linkage disequilibrium (Eldon & Wakeley 2008; Birkner *et al.* 2012), will depend on the frequency of recombination in a given parasite species . However, a future challenge is to distinguish such signatures of natural selection (positive or negative) from the genome wide variance in topology and length under occurring MMC genealogies. The large variance in possible MMC genealogy originates from random neutral processes such as sweepstake reproduction, recurrent bottlenecks and alternate asexual and sexual phases as discussed above.

**Conclusions**

We highlight here several species and life history traits which promote variance in offspring production and thus depart from the assumptions of the classic WF model and Kingman coalescent. Analysing genomic data in such species may thus lead to potential source of error, misinterpretation of past demography and false detection of genes under selection. We advocate here that this burst of new MMC coalescent models represents a chance to 1) improve model based inference in many bacterial, fungal, plant and animal species, and 2) to connect population genetics models with life-history traits. However, efforts remain to be made to test the relevance of these various models ($\Lambda$-coalescent, beta-coalescent, Bolthausen-Snitzman, $\Xi$-coalescent) compared to existing extensions of the Kingman model. Finally, coalescent theory does not only allow us to draw inference from polymorphism data about past demography and the action of selection in the genome, it also provides a theoretical framework to develop new statistical methods and a renewed understanding of genome data in numerous bacteria, fungi, viruses, marine organisms or plant species with peculiar life-cycles.

**Acknowledgments**

The authors are indebted to Matthias Birkner, Bjarki Eldon, Fabian Freund for their patience in answering questions, for useful discussions and comments on the manuscript. We thank M. Birkner for sharing R codes for preparing Figure 2, and Jerome Enjalbert for inspiration for Figure 3B. AT acknowledges support the German Federal Ministry of Education and Research (BMBF) within the AgroClustEr "Synbreed – Synergistic plant and animal breeding" (FKZ: 0315528I).

**Figure legends**

*Figure 1: Schematic genealogies for three types of coalescent models:* A) Genealogy based on the Kingman's coalescent with at most only one coalescent event per generation between two lineages. B) Genealogy with a maximum of one coalescent event per generation involving two or more lineages applicable to the Λ-coalescent, Beta-coalescent and ψ-coalescent. C) The general genealogy with simultaneous multiple coalescent of two or more lineages per generation as found in the Ξ-coalescent and Bolthausen-Snitzman model. Note that the length of genealogies may vary considerably between MMC trees and is here not indicative.

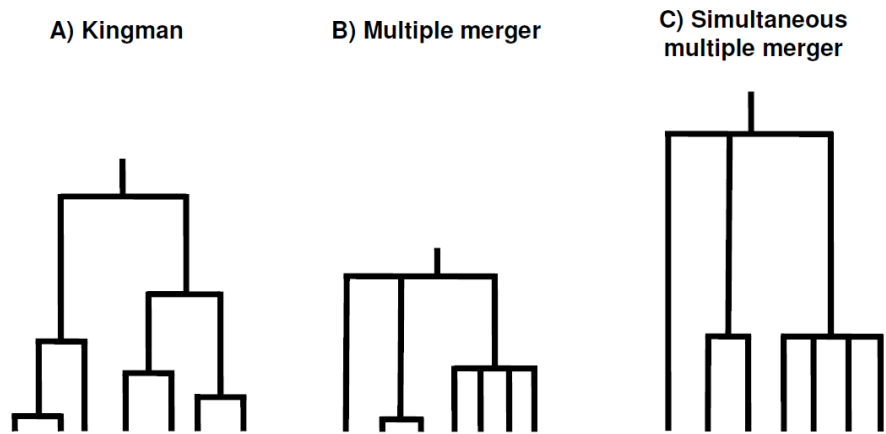

*Figure 2: Site frequency-spectrum at one locus with sample size n = 20 for different coalescent models.* Three models are compared: the neutral model of Kingman with constant population size (black bars), the beta-coalescent (blue bars), and the Kingman coalescent with step-wise expanding population (red bars). A) SFS for beta-coalescent with parameter α = 0.8 (blue) and population expansion with a growth factor of 40 (red). B) SFS for beta-coalescent with parameter α = 1.1 (blue) and population expansion with a growth factor of 10 (red). C) SFS for beta-coalescent with parameter α = 1.7 (blue) and population expansion with a growth factor of 2 (red).

The average SFS over 10,000 replicates is shown, assuming a population mutation rate $\theta$ = 30 for the Kingman models, and $r = 10$ for the Beta-coalescent model. The Beta-coalescent simulations are computed using the recursion (R-code) from Birkner et al. (2008).

A)

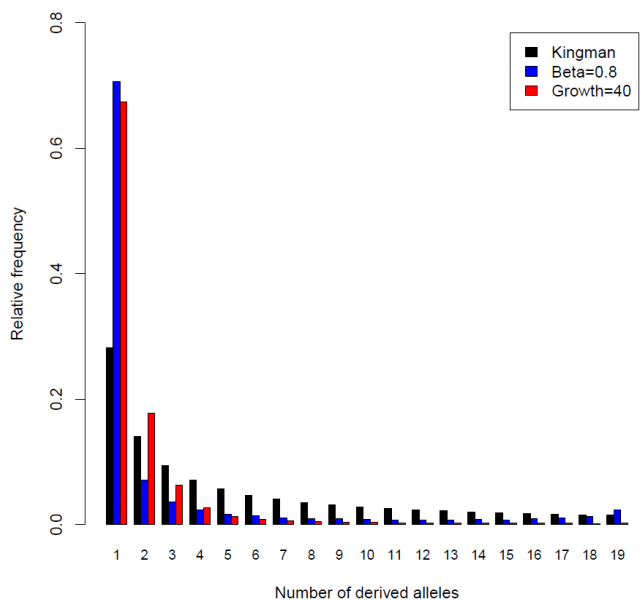

B)

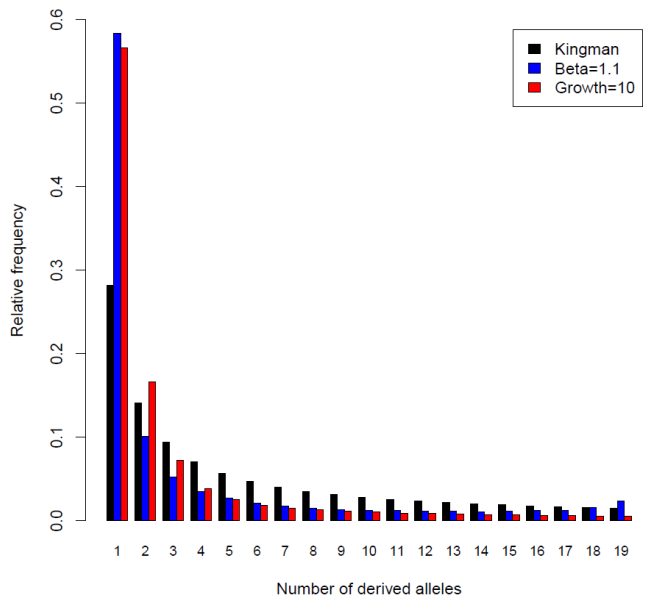

C)

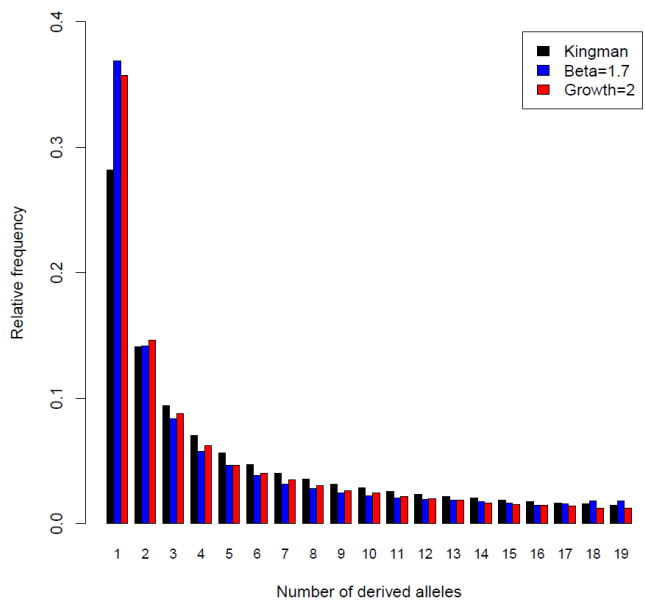

*Figure 3: Schematic representation of key life cycles and life history traits giving rise to multiple merger coalescent models.* A) Model of sweepstake reproduction common to many in marine organisms with N being the adult population size, which is also the carrying capacity of the population. Variance in offspring production is very large, and the production of eggs and juveniles largely exceeds the carrying capacity. Density-dependent regulation thus maintains the population at size *N*. B) Demographic events drive random neutral evolution in crop pathogens over three years with seasonality (winter = W, summer = S). As the host plants grow during the year, the pathogen population grows. Harvest decreases dramatically the availability of hosts, as pathogen can only survive then on volunteer crops or wild relatives, generating regular periodic bottlenecks in pathogen density (y-axis). C) Events in the parasite population: neutral stochastic events during inter-host transmission and selective events during intra-host dynamics. Host individuals are denoted by the green solid rectangles, while parasite particles are the small filled circles. Adaptation in the parasite population is continuous in maintaining the fittest individuals within each infected host (green dotted arrow), while recurrent bottlenecks occur during inter-host transmission (black dotted arrows).

## A) Sweepstake reproduction

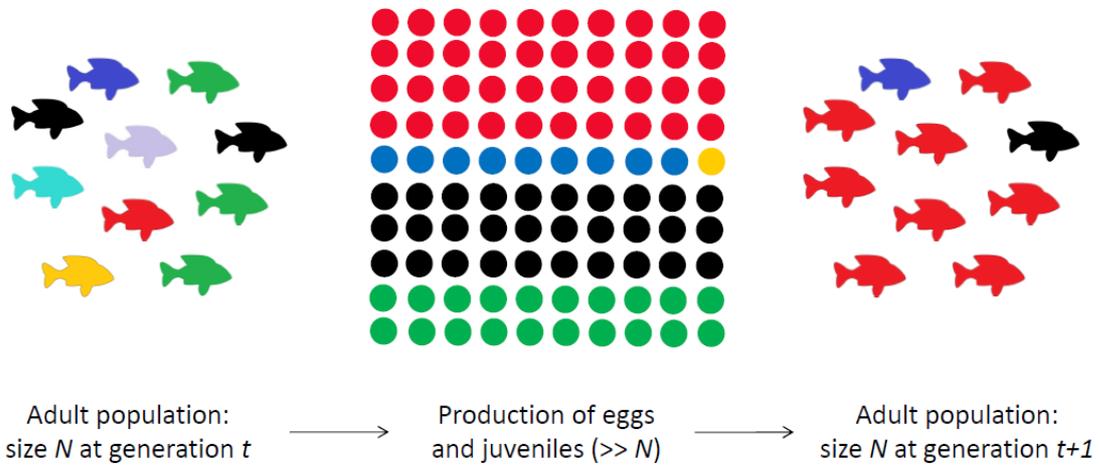

Adult population: size *N* at generation *t* → Production of eggs and juveniles (>> *N*) → Adult population: size *N* at generation *t+1*

## B) Recurrent bottlenecks in crop pathogen populations

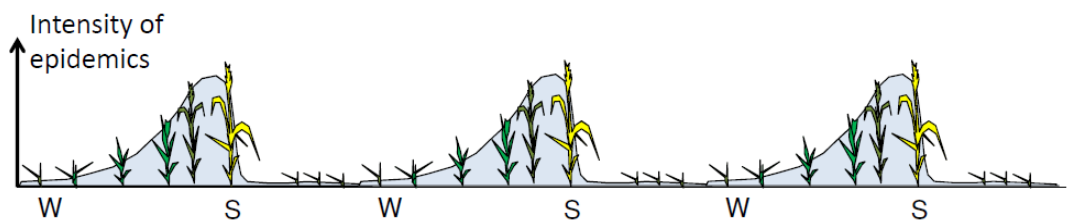

## C) Selection and bottlenecks in rapidly adapting parasite populations

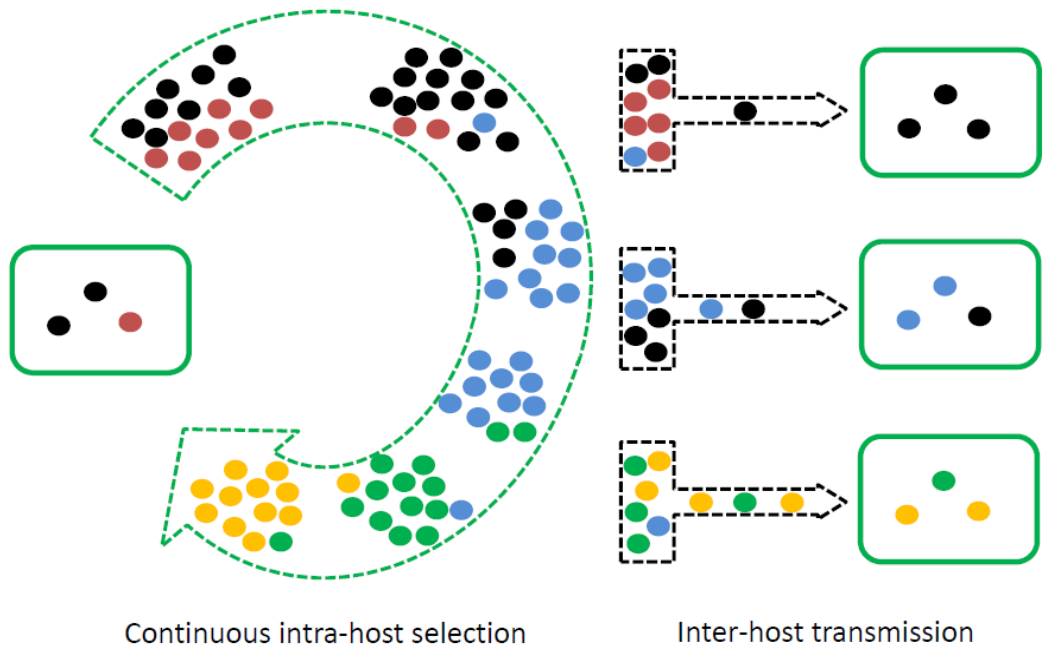

Continuous intra-host selection          Inter-host transmission